\documentclass[twocolumn,prl,aps,showpacs]{revtex4}

\usepackage{graphicx}%

\usepackage{dcolumn}
\usepackage{amsmath}
\usepackage[caption=false]{subfig}

\newcommand{\ed}{\end{document}}

\newcommand{\ice}[1]{\relax}

\newcommand{\re}[1]{(\ref{#1})}

\newcommand{\as}{a_s}

\newcommand{\beq}{\begin{equation}}

\newcommand{\ba}{\begin{array}}

\renewcommand{\arraystretch}{1.7}

\newcommand{\ea}{\end{array}}

\newcommand{\eeq}{\end{equation}}

\newcommand{\bea}{\begin{eqnarray}}

\newcommand{\eea}{\end{eqnarray}}
\newcommand{\g}{\gamma}

\newcommand{\ovl}{\overline}

\newcommand{\nnb}{\nonumber}
\newcommand{\sbz}{  }

\def\bbuildrel#1_#2^#3%
{\mathrel{\mathop{\kern 0pt#1}\limits_{#2}^{#3}}}

\begin{document}

\title{
Five-Loop Running  of the  QCD coupling constant\footnote{
In memoriam  Dmitry Vasil'evich Shirkov, 1928-2016}
 }     

\author{P.~A.~Baikov}
\affiliation{
Skobeltsyn Institute of Nuclear Physics, Lomonosov Moscow State University\\
1(2), Leninskie gory, Moscow 119991, Russian Federation
        }

\author{K.~G.~Chetyrkin}
\author{J.~H.~K\"uhn}
\affiliation{Institut f\"ur Theoretische Teilchenphysik, Karlsruhe
  Institute of Technology (KIT), Germany}

\begin{abstract}
\noindent

We analytically compute the five-loop term in the  beta function
which  governs the running of $\alpha_s$ --- the quark-gluon coupling constant in QCD.
The new term leads to  a reduction
of the theory uncertainty in $\alpha_s$ taken at the Z-boson scale  as
extracted from the $\tau$-lepton decays as well as to  new, improved  by one
more order of perturbation theory, predictions for the effective coupling
constants of the Standard Model Higgs boson  to gluons and
for its total decay rate to the quark-antiquark pairs.

\end{abstract}

\pacs{12.38.-t 12.38.Bx; Preprint  TTP16-026}

\maketitle


Asymptotic freedom, manifest by a decreasing coupling with increasing energy,
can be considered as the basic prediction of nonabelian gauge theories and was
crucial for establishing Quantum Chromodynamics (QCD) as the theory of strong
interactions \cite{Gross:1973id,Politzer:1973fx}.  The dominant, leading order
prediction was quickly followed by the corresponding two-loop
\cite{Caswell:1974gg,Jones:1974mm} and three-loop
\cite{Tarasov:1980au,Larin:1993tp} results. The next, four-loop calculation
was performed almost twenty years later \cite{vanRitbergen:1997va} and
confirmed in \cite{Czakon:2004bu}.  These results have moved the theory from
qualitative agreement with experiment, as observed on the basis of the early
results, to precise quantitative predictions, valid over a wide kinematic
range, from $\tau$-lepton decays up to LHC results.

Although the agreement between theory predictions and experimental
results is impressive already now, it is tempting to push the theory
prediction as high as possible. On the one hand one may expect an even
better agreement between theory and experiment. On the other hand
it is of theoretical interest to push gradually into the region where
individual terms of the series might start to increase, thus
demonstrating the asymptotic divergence of the perturbative series.
At a more modest level we note that  predictions for the five-loop term
that can be found in the literature are based on a variety of methods
and  exhibit for some cases quite  a dramatic  variation of 
the size of  the  term (we will give more    details later).

There are, of course, a number of phenomenological applications of the
five-loop result, which will be discussed in this paper. On the one hand
there is the relation between $Z$-boson and  $\tau$-lepton decay rates
into hadrons, which involves the strong coupling at two vastly different
scales. On the other hand we will discuss the Higgs boson decay rate
into bottom quarks and into gluons, which are sensitive to the five-loop
running of the QCD coupling.

Let us start with the definition of the beta function
\begin{equation}
\beta(a_s)=\mu^2\frac{d}{d\mu^2}a_s(\mu)=
-\sum_{i\ge 0} \beta_i a_s^{i+2}
\end{equation}
which describes the running of the quark-gluon  coupling $a_s\equiv \alpha_s/\pi$ as
a function of the normalization scale $\mu$ within the renormalization group
approach \cite{Stueckelberg53,GellMann:1954fq,Bogolyubov:1956gh}.
\ice{ with the first term 
$\beta_0= (33 -2 \, n_f)/12 $, with $n_f$  being the number of active  quark 
flavors. 
}

Using the same theoretical tools as   in the calculations of  \cite{Baikov:2008jh} and \cite{Baikov:2014qja}
we have computed  the   QCD  $\beta$-function in five-loop order with the result
\begin{eqnarray}
\renewcommand{\arraystretch}{ 1.3}
 \beta_0 & = & \frac{1}{4}\Biggl\{ 11 - \frac{2}{3} n_f,  \Biggr\}, \ \
\beta_1    =  \frac{1}{4^2}\Biggl\{ 102 - \frac{38}{3} n_f\Biggr\}, 
\\
 \beta_2   &=&   \frac{1}{4^3}\Biggl\{\frac{2857}{2} - \frac{5033}{18} n_f + \frac{325}{54}
  n_f^2\Biggr\}, 
 \\
 \beta_3 & = &  \frac{1}{4^4}\Biggl\{ 
\frac{149753}{6} + 3564 { \zeta_3} 
        - \left[ \frac{1078361}{162} + \frac{6508}{27} { { \zeta_3}} \right] n_f
   \\ & & \hspace{15mm}
       + \left[ \frac{50065}{162} + \frac{6472}{81} { \zeta_3} \right] n_f^2
       +  \frac{1093}{729}  n_f^3\Biggr\}
{},
\nonumber
\ice{
\end{eqnarray}
\begin{eqnarray}
}
\\
&{}&\hspace{-4.5mm}\beta_{4} =  \frac{1}{4^5}\,\Biggl\{
\frac{8157455}{16} 
+\frac{621885}{2}  \sbz \zeta_{3}
-\frac{88209}{2}  \sbz \zeta_{4}
-288090  \sbz \zeta_{5}
\nonumber\\
&{+}& \, n_f
\,\, 
\left[
-\frac{336460813}{1944} 
-\frac{4811164}{81}  \sbz \zeta_{3}
\right.
\nnb
\\
&{}& \hspace{40mm}
\left.
+\frac{33935}{6}  \sbz \zeta_{4}
+\frac{1358995}{27}  \sbz \zeta_{5}
\right]
\nonumber\\
&{+}& \, n_f^2
\,\,
\left[
\frac{25960913}{1944} 
+\frac{698531}{81}  \sbz \zeta_{3}
-\frac{10526}{9}  \sbz \zeta_{4}
-\frac{381760}{81}  \sbz \zeta_{5}
\right]
\nonumber\\
&{+}& \, n_f^3
\,\,
\left[
-\frac{630559}{5832} 
-\frac{48722}{243}  \sbz \zeta_{3}
+\frac{1618}{27}  \sbz \zeta_{4}
+\frac{460}{9}  \sbz \zeta_{5}
\right]
\nonumber
\\
&{+}&  \, n_f^4\,\,
\left[
\frac{1205}{2916} 
-\frac{152}{81}  \sbz \zeta_{3}
\right]
\Biggr\}
\label{beta_5l}
{},
\end{eqnarray}
where $n_f$ denotes  the number of active  quark 
flavors. 
As expected from the three and four-loop results, the higher
transcendentalities $\zeta_6$ and $\zeta_7$ that could be present at five-loop
order \footnote{For a general analysis of the issue,  see  \cite{Baikov:2010hf}.}, 
are actually absent.  Note that the contribution in $\beta_4$ that is
leading in $n_f$ (proportional to $n_f^4$) was computed long ago with a very
different technique \cite{Gracey:1996he} for a generic gauge group. 
For the physical case of  $SU(3)$ we find full agreement.

In  numerical form the coefficients $\beta_0 - \beta_4$ read
\begin{eqnarray}
 \beta_0 & \approx &  2.75 - 0.166667 \, n_f \nonumber ,\\
 \beta_1 & \approx & 6.375 - 0.791667 \, n_f\nonumber ,\\
 \beta_2 & \approx & 22.3203 - 4.36892 \, n_f + 0.0940394 \, n_f^2 \nonumber, \\
 \beta_3 & \approx & 114.23 - 27.1339 \, n_f \nnb
        \\     
&{}& \hspace{12mm}         +\, 1.58238 \, n_f^2 + 0.0058567 \, n_f^3 \nnb
         ,\\
 \beta_4 & \approx & 524.56 - 181.8 \, n_f + 17.16 \, n_f^2 
           \nnb
\\ &{}& \hspace{1cm}       - \,\,0.22586 \, n_f^3 -  0.0017993 \, n_f^4
{}.
\label{beta4N}
\end{eqnarray}

Numerically the coefficients are surprisingly small.
For example,  for the  particular cases of $n_f =3,4,5 $ and $6$
we get:
\bea
\nnb \ovl{\beta}(n_f=3) &=&
1 +1.78 \,a_s+4.47 \,a_s^2+20.99 \,a_s^3+56.59 \,a_s^4,
\\ \nnb \ovl{\beta}(n_f=4) &=&
1 +1.54 \,a_s+3.05  \,a_s^2+15.07  \,a_s^3+27.33 \,a_s^4,
\\ \nnb \ovl{\beta}(n_f=5) &=& 
1 +1.26 \,a_s+1.47 \,a_s^2+9.83 \,a_s^3+7.88 \,a_s^4,
\\    \ovl{\beta}(n_f=6) &=& 
\nnb
1 +0.93  \,a_s-0.29  \,a_s^2+5.52  \,a_s^3+0.15 \,a_s^4
{},
\eea
where 
$
\ovl{\beta} \equiv \frac{\beta(a_s)}{-\beta_0 \as^2} 
= 1 + \sum_{i\ge 1}\bar{\beta_i} a_s^i
{}.
$
A very modest growth of the coefficients is observed  and the (apparent) convergence
is better than one would expect from comparison with other examples.

It is  instructive to compare $\beta_4$ as shown in eq.~(\ref{beta4N}) with a (20 years old!) prediction
based on the so-called method of the Asymptotic Pade Approximant (APAP)   
from \cite{Ellis:1997sb} (the boxed term was used as input):
\[  
\beta_4^{APAP}   =  740 - 213\, n_f + 20\, n_f^2  -0.0486\, n_f^3   -\fbox{$
  0.0017993 \,n_f^4$} 
{}.
\]
Unfortunately, this  strikingly  good agreement for all powers of $n_f$ except for $n_f^3$ term
does { not always } survive for  fixed values of $n_f$ due to huge  cancellations between
contributions proportional to  different   powers of $n_f$ (see  Table \ref{table1} below).
\vspace{-5mm}
\begingroup
\squeezetable
\begin{table}[h!]
\caption{
\label{table1} 
Comparison  of the exact  results for $\beta_4$ with the  predictions based on
APAP  for different values of $n_f$.
 }
\begin{ruledtabular}
\begin{tabular}{c c c  c  c   c c c  }
    $n_f$               & 0  &  1 &  2      &  3       & 4     &  5  &     6   \\ \hline
    $\beta_4^{\rm exact}$ & 525 &  360 &  228   &  127   &  57    &  15  &    0.27\\    \hline
    $\beta_4^{\rm APAP}$ &  741 &  548&   395  &   281    &  205   &  169 &
    170  \\ 
  \end{tabular}
\end{ruledtabular}
\end{table}
\endgroup

At this point it may be useful to present  
the impact  of the five-loop term on the running of the strong  coupling  from
low energies, say $\mu=M_{\tau}$, up to the high energy region $\mu=M_H$, 
by comparing  the predictions  based on three and four versus five-loop
results \footnote{
For all practical examples  in this  paper  we  have  used an extended version
of the package RunDec  \cite{Chetyrkin:2000yt}.}.
We start from the scale of  $M_{\tau}$ with $\alpha_s^{(3)}(M_{\tau})= 0.33$
  (as given in \cite{pdg2014})   
 and evolve the coupling up  to 3 GeV.
At this point  the four-loop matching from  3 to 4 flavours  is  performed. 
The  strong coupling  now runs up to $\mu=10$  GeV and, at this point, the
number of  active  quark flavours  is 
switched  from  the 4 to 5. Subsequently,  the strong coupling  runs again 
up to $M_Z$ and,  finally,  up to the Higgs mass  $M_H =125$ GeV.  The
relevant  values of $\alpha_s$ are  listed  in Table \ref{table2}.
The combined uncertainty in $\alpha_s^{(5)}(M_Z)$  induced by  running and 
matching can be conservatively estimated by the shift in  $\alpha_s^{(5)}(M_Z)$
produced  by the use of five-loop  running (and, consequently) four-loop matching 
instead of four-loop running  (and three-loop matching). It amounts to a 
minute $8\cdot 10^{-5}$ which is by a factor of three less than the similar
shift made by the use of four-loop running instead of the the three-loop one 
(see Table \ref{table2}). Note that the final  value of $\alpha_s^{(5)}(M_Z)$ 
which follows from  $\alpha_s^{(3)}(M_{\tau})$  is in   remarkably
good agreement with the fit to electroweak precision data (collected in $Z$
boson decays), namely (\cite{pdg2014}):
\beq
\alpha_s^{(5)}(M_Z)  = 0.1197 \pm 0.0028
{}.
\eeq
 \vspace{-5mm} 
\begin{table}[h!]
{ \caption{
\label{table2} 
Running of $\alpha_s$ from $\mu=M_{\tau}$ to $\mu=M_H$. 
For the threshold values of $c$ and $b$ heavy quarks we have chosen
\cite{Chetyrkin:2009fv,Chetyrkin:2015mxa}
 $m_c(3\,\mbox{ GeV}) = 0.986  \,\mbox{ GeV}$ and  $m_b(10\,\mbox{ GeV}) = 3.160 \,\mbox{ GeV}$
respectively. 
 }
}
\begin{center}
\begin{tabular}{|c|c|c|c|}
\hline
 \# of loops  & $\alpha_s^{(3)}(M_{\tau})$  & $\alpha_s^{(5)}(M_Z)$ & $\alpha_s^{(5)}(M_H)$
 \\
 3  &$ 0.33  \pm 0.014$  & $0.1195  \pm 0.0015$ & $0.1140 \pm 0.0015$\\
 4  &$ 0.33 \pm 0.014$   & $0.1197  \pm 0.0015$ & $0.1142  \pm 0.0015$ \\
 5  &$ 0.33 \pm 0.014$   & $0.1198 \pm 0.0015$  & $0.1143   \pm 0.0015$ \\
\hline
\end{tabular}
\end{center}
\end{table}
\vspace{-1mm}
\ice{
mc3GeV   =  0.986 + pm*.013/.pm->0;

mc10GeV  =  0.5 + pm*.013/.pm->0;

mb10GeV  =  3.610 + pm* 0.016/.pm->0  - 0*(alsMz -  0.1189)/.002;
}
As anticipated in \cite{Baikov:2014qja}, the
running of $m_b$ from low energies, say 10~GeV, is affected by the
five-loop term, which in turn, slightly modifies the Higgs boson decay
rate into a quark pair. This rate is given by
\begin{equation}
\Gamma(H\to f\bar f)=
\frac{G_F M_H}{4\sqrt{2}\pi} m_f^2(\mu) R^S(s=m_H^2,\mu)
{},
\end{equation}
where $\mu$ is the normalization scale and $R^S$ the spectral density of
the scalar correlator, known to $\alpha_s^4$ from \cite{Baikov:2005rw}
\ice{

RSm0as4/.LMS->0/.nl-> 5

                                    2             3             4
Out[5]= 1. + 5.66667 as + 29.1467 as  + 41.7576 as  - 825.747 as

                                     2               3               4
Out[6]= 1. + 0.206169 h + 0.0385818 h  + 0.00201106 h  - 0.00144688 h

0.1143/Pi*h

Out[7]= 0.0363828 h
}
\vspace{-1mm}
\bea
&{}& R^{S}(s = M_H^2,\mu=M_H)  
\nnb
\\
&{=}& 1 + 5.667\,  a_s+ 29.147 \, a_s^2  +
  41.758 \,a_s^3 \,  {- 825.7}\,a_s^4
,
\nnb
\\
&=& 
1 + 0.2062  + 0.0386   + 0.0020  {-0.00145}
\label{RS_as4_nl5}
{},
\eea
where we set
$a_s(M_H)=\alpha_s(M_H)/\pi=0.1143/\pi=0.0364$ and $R^S$  is evaluated for the
Higgs mass value \mbox{$M_H =125 \,\mbox{GeV}$}.
For the running of the $b$ quark mass the corresponding input is taken from a
relatively low scale and has to be evolved up to $M_H$. The shift from
the five-loop term is then given by
\begin{equation}
\frac{\delta m_b^2(M_H)}{m_b^2(M_H)} = -1\cdot\, 10^{-4}
\end{equation}
which at present and in the foreseeable future is negligible. 
We want to stress here that the effect due to the ${\cal O}(\alpha_s^4)$ term in 
(\ref{RS_as4_nl5}) are formally  of the same order as the one induced by 
the five-loop running of $m_b$.

\ice{
(* 
In contrast, a value of $\beta_4$ close to  135, as predicted in  \cite{Elias:1998bi},
would lead to a sizable shift of the prediction.
 *)
}

Another application of our result for the $\beta$-function is 
the determination of the effective Higgs-gluon-gluon coupling.  
In the heavy top limit the  Higgs boson  couples directly with gluons via
the effective Lagrangian of the form
\cite{Wilczek:1977zn,Shifman:1979eb,Inami:1982xt,Kniehl:1995tn}
\beq
{\cal L}_{\mathrm eff} = -2^{1/4}G_F^{1/2} H C_1(\mu^2/m_t^2,a_s(\mu)) \, G^a_{\nu\rho} G^a_{\nu\rho}
{}.
\eeq
The effective coupling constant $C_1(\mu^2/m_t^2,a_s(\mu))$ appears as a
common factor in two quantities important for Higgs physics
processes, namely,   Higgs decay into  gluons (one of the  main decay channels for  the Standard Model Higgs boson)
and  Higgs production  via  the gluon fusion (the main Higgs  production mode on LHC).
It is expressible through massive tadpoles and was computed at four loops in
1997 \cite{Chetyrkin:1997un} (long before the direct calculation of four-loop generic massive
tadpoles started to be technically feasible). This happened to be possible due
to a   low energy theorem  (exact in all orders) \cite{Chetyrkin:1997un}
\beq
C_1 = -\frac{1}{2} m_t^2 \frac{\partial}{\partial  m_t^2} \ln \zeta_g^2, 
\ \ \alpha_s'(\mu) = \zeta_g^2(\mu^2/m_t^2,\alpha_s(\mu))  \, \alpha_s(\mu) 
\label{C1}
\eeq
which connects $C_1$ with the corresponding ``decoupling'' constant $\zeta_g$ for
$\alpha_s$.  The appearance of the derivative $\frac{\partial }{\partial
m_t^2}$ means that the most complicated (that is constant) part of $\zeta_g^2$
does not contribute to $C_1$, so that one could use the corresponding RG
equation to find logs at next loop order ({\em provided we know the 
$\beta$-function at the same increased loop order!}). 
  
Since  the decoupling constant  is  known at four loops from \cite{Chetyrkin:2005ia,Schroder:2005hy}
we can  now  use \re{C1} and  \re{beta_5l} to extend the known  four-loop result to one  more loop:
\begin{align}
C_1 = -\frac{1}{12} a_s \, \Bigl( 1 + 2.750\ a_s + 6.306\, a_s^2 + &  \  4.794\, a_s^3
\label{C1}
\\
+  & \ 41.447 \, a_s^4 \Bigr)
{}.
\nnb
\end{align}
In this expression $a_s = \alpha_s^{(6)}(\mu_t)/\pi$,  with  $\mu_t$ being a
scale-invariant top quark mass defined as $\mu_t =m_t(\mu_t)$.
Note that  the contribution due to $\beta_4$  to the last coefficient (boxed below)
is significant, namely, 
\beq
 41.447  =  -47.611  + \fbox{89.058}.
\label{C1:decomposition}
\eeq
\ice{
12.427 + 89.058

Out[2]= 101.485
}

As another application let us  mention the connection  with  the
renormalization  group invariant (RGI) mass:
\beq
{m}^{RGI} \equiv m(\mu_0)/{c(a_s(\mu_0))}
{},
\eeq
with 
\beq
\frac{m(\mu)}{m(\mu_0)} = \frac{c(a_s(\mu))}{c(a_s(\mu_0))}, \  \ \ 
c(x) = \mathrm{exp}\Biggl\{ \int {d x'} \frac{\g_m(x')}{\beta(x')} \Biggr\} 
{},
\label{cfun:1}
\eeq
which could be determined in lattice calculations.  The function $c(x)$ does
depend not only on the quark mass anomalous dimension $\gamma_m$ (known from
\cite{Vermaseren:1997fq,Baikov:2014qja}) but also on the $\beta$-function. In the five-loop
approximation we get (for  a typical for lattice simulations value of $n_f=3$)
\bea
c(x)\bbuildrel{=\!=\!=}_{n_f = 3}^{}  x^{4/9}  \Bigl(
 1  + 0.8950 \,x + 1.3714 \,x^2  + 1.9517 \,x^3
\nnb
\\ 
 + (15.6982 -  0.11111\,\bar{\beta_4} = 9.411) \,x^4   
\Bigr),\hspace*{2mm}
\label{ms_rgi}
\eea
with $\bar{\beta}_4 = \beta_4/\beta_0 = 56.59$.

The precise knowledge of the function $c(x)$ (which is a scheme dependent quantity) is
required in order to find   the  mass of the strange quark in a well-defined
renormalization scheme  (usually the $\ovl{\mbox{MS}} $-one) from
$m_s^{RGI}$ measured with lattice
simulations at  very  high energies around 100 GeV \cite{Sommer:2015kza}. 
With a  typical  value of $\alpha_s(2 \, \mbox{GeV} )/\pi= 0.1$ we find that the series 
\re{ms_rgi} shows   quite  good convergence.
In contrast, a value of $\beta_4$ as large as $-2000$  as estimated in \cite{Elias:1998bi}
would lead to a significantly less  stable series.

Summary: The exact result for  the  five-loop term of the QCD $\beta$-function 
allows to relate  the  strong coupling constant $\alpha_s$, as determined  with
NLO${}^3$ accuracy at low energies, say $M_{\tau}$ with  the strong coupling
as evaluated at high scales, say $M_Z$  or $M_H$. Including the exact
five-loop term  has little influence  on the central value of the  prediction,
a consequence  of partial  cancellations between various  contributions
from  matching and running. However,  the five-loop  result leads  to a
considerable  further  reduction  of the  theory  uncertainty and allows
to combine values from low and high energies of appropriate order. It also
should be  useful in the elimination of  the renormalization scheme and scale ambiguities in perturbative QCD
within the framework of  The Principle of Maximum Conformality
and Commensurate Scale Relations \cite{Brodsky:2013vpa} or, closely related,
the sequential extended BLM approach \cite{Mikhailov:2004iq,Kataev:2014zwa}.

We want now to add here some technical details about our calculation.  To
evaluate the $\beta$-function we need to evaluate the following three
renormalization constants (RC's) in five-loop order: $Z^{ccg}_1$ for the
ghost-ghost-gluon vertex, $Z_3^c$ for the inverted ghost propagator and $Z_3$
for the inverted gluon propagator.  The total number of five-loop diagrams
contributing to the RC's (as generated by QGRAF \cite{QGRAF}) amounts to about
one  and a  half  million (1.5$\,\cdot 10^6$), with the gluon wave-function
$Z_3$ (around 3$\,\cdot 10^5$ diagrams) being most complicated one.  Every
power of  $n_f$ in \re{beta_5l} was computed separately with the help of the
FORM \cite{Vermaseren:2000nd,Steinhauser:2015wqa} program BAICER,
implementing the algorithm of works
\cite{Baikov:2005nv,Baikov:1996rk,Baikov:2007zza}.

With a typical set-up of 15-20 workstations (with 8 cores each) running a
thread-based version of FORM \cite{Tentyukov:2007mu} the calculation of two
first subproblems ($n_f^4$ and $n_f^3$) took together about a couple of weeks,
while the remaining three most complicated pieces (proportional to $n_f^2$,
$n_f^1$ and $n_f^0$ correspondingly) required up to 7 months of running time
for every particular $n_f$-slice.

The continued running of our calculations at such computer and time scales
would be virtually impossible without the effective support of our computer
administration, in particular, Alexander Hasselhuhn, Jens Hoff, David Kunz,
Peter Marquard and Matthias Steinhauser, to whom all we express our sincere
thanks.

We are gratefull to Michael Spira for the carefull reading of the first
version of the paper and letting us know about numerical errors (fixed in the
current version) in eqs.~(\ref{C1},\ref{C1:decomposition}).

The work by K.~Chetykin and J.~H.~K\"uhn was supported by the Deutsche
Forschungsgemeinschaft through CH1479/1-1.  The work of P.A.Baikov is
supported in part by grant NSh-7989.2016.2 of the President of Russian
Federation.

After our calculations had been finished we have been informed that the
subleading in $n_f$ term in the coefficient $\beta_4$ (proportional to $n_f^3$
in eq.~(\ref{beta_5l})) has been confirmed and even extended for the case of a
general gauge group in \cite{Luthe:2016ima}. The authors have used a radically
different method which expresses the $\beta$-function in terms of completely
massive vacuum diagrams.

The work is dedicated to the memory of one of the founders of the
renormalization group method---Dmitry Vasil'evich Shirkov, 1928-2016.

\ed

\bibliographystyle{apsrev}
\bibliography{acat,brodsky,higgs_as5,chet,gestimations,chet_full,gracey,schroder,LiteraturSM,baikov,beta5,lattice,rg}

\end{document}